\documentclass[10pt,a4paper]{article}
\usepackage{amsmath,amssymb}
\begin{document}

\title{Order Topology and Frink Ideal Topology of Effect Algebras\thanks {This project is supported by Natural Science Foundation of China (10771191 and
10471124) and Natural Science Foundation of Zhejiang Province of
China (Y6090105).}}
\author{Qiang Lei\thanks{Department of Mathematics, Harbin Institute of Technology, Harbin, China; e-mail: leiqiang@hit.edu.cn}\and
Junde Wu\thanks{Department of Mathematics, Zhejiang University,
Hangzhou, China; e-mail: wjd@zju.edu.cn}\and Ronglu
Li\thanks{Department of Mathematics, Harbin Institute of Technology,
Harbin, China. }}
\date{}
\maketitle

\noindent {\bf Abstract}. In this paper, the following results are
proved: (1) $ \ $ If $E$ is a complete atomic lattice effect
algebra, then $E$ is (o)-continuous iff $E$ is order-topological iff
$E$ is totally order-disconnected iff $E$ is algebraic. (2) $ \ $ If
$E$ is a complete atomic distributive lattice effect algebra, then
its Frink ideal topology $\tau_{id}$ is Hausdorff topology and
$\tau_{id}$ is finer than its order topology $\tau_{o}$, and
$\tau_{id}=\tau_o$ iff $1$ is finite iff every element of $E$ is
finite iff $\tau_{id}$ and $\tau_o$ are both discrete topologies.
(3) $ \ $  If $E$ is a complete (o)-continuous lattice effect
algebra and the operation $\oplus$ is order topology $\tau_o$
continuous, then its order topology $\tau_{o}$ is Hausdorff
topology. (4) $ \ $ If $E$ is a (o)-continuous complete atomic
lattice effect algebra, then $\oplus$ is order topology continuous.


\vskip 0.1 in


\vskip 0.2 in

{\bf  1. Introduction }

\vskip 0.2 in

Effect algebra is an important model in studying the unsharp quantum
logic theory, it is also an important carrier of quantum states and
quantum measures ([1]). As an important tool of studying the quantum
states and quantum measures, the topological structures of effect
algebras not only can help us to describe the convergence properties
of quantum states and quantum measures, but also can help us to
characterize some algebra properties of effect algebras. This paper
contributes to the understanding of the topological properties and
algebraic properties of effect algebras, it both promotes some
classical results, for example, Theorem 2.1, and obtain several new
interesting conclusions, for example, Theorem 3.1, Theorem 4.1 and
Theorem 4.3, etc. Now, we show them in details in the following
three sections. 

\vskip 0.1 in

The structure $(E, \oplus, 0, 1)$ is said to be an effect algebra if
0, 1 are two distinguished elements of $E$ and $\oplus$ is a
partially defined binary operation on $E$ which satisfies the
following conditions for any $a, b, c\in E$ ([1]):

\vskip 0.1 in

(1) $ \ $ If $a\oplus b$ is defined, then $b\oplus a$ is defined and
$a\oplus b=b\oplus a$.

(2) $ \ $ If $a\oplus b$ and $(a\oplus b)\oplus c$ are defined, then
$b\oplus c$ and $a\oplus (b\oplus c)$ are defined and $(a\oplus
b)\oplus c=a\oplus (b\oplus c)$.

(3) $ \ $ For each $a\in E$ there exists a unique $b\in E$ such that
$a\oplus b$ is defined and $a\oplus b=1$.

(4) $ \ $ If $a\oplus 1$ is defined, then $a=0$.

\vskip 0.1 in

The effect algebra $(E, \oplus, 0, 1)$ is often denoted by $E$. For
every $a\in E$, we denote the unique $b$ in condition (3) by $a'$
and call it the orthosupplement of $a$. The sense is that if $a$
presents a proposition, then $a'$ corresponds to its negation. The
operation $\oplus$ of an effect algebra $E$ can induce a new partial
operation $\ominus$ and a partial order $\leq$ as follows: $a\ominus
b$ is defined iff there exists $c\in E$ such that $b\oplus c$ is
defined and $b\oplus c=a$, in which case we denote $c$ by $a\ominus
b$; $d\leq e$ iff there exists $f\in E$ such that $d\oplus f$ is
defined and $d\oplus f=e$. This showed that every effect algebra is
a partial order set. If $(E, \leq)$ is a lattice, then $E$ is called
a lattice effect algebra, similarly, we can define the complete
lattice effect algebras. If a lattice effect algebra is a
distributive lattice, then it is called a distributive lattice
effect algebra. For more details on effect algebras, for example,
$a\oplus b$ is defined iff $a\leq b'$, we refer to [1].

\vskip 0.1 in

Let $E$ be an effect algebra and $a,b\in E$ with $a\leq b$, denote
$[a,b]=\{p\in E|a\leq p\leq b\}$. A nonzero element $a\in E$ is said
to be an atom of $E$ if $[0,a]=\{0,a\}$, $E$ is said to be atomic
iff for every nonzero element $p\in E$, there is an atom $a\in E$
such that $a\leq p$ ([1]).

\vskip 0.1 in

An element $x\in E$ is said to be a sharp element of $E$ if $x\wedge
x'=0$, that is, proposition $x$ and its negation $x'$ have no
overlaps.

\vskip 0.2 in

{\bf  2. The order-continuity and order-topological effect algebras}

\vskip 0.2 in

Assume that $P$ is a partial order set and
$(a_\alpha)_{\alpha\in\varepsilon} $ is a net of $P$. If for any
$\alpha,\beta \in \varepsilon$, when $\alpha\preceq\beta$,
$a_\alpha\leq a_\beta$, then we denote $a_\alpha\uparrow$, moreover,
if $a=\bigvee\{a_\alpha|\alpha\in \varepsilon\}$, then we denote
$a_\alpha\uparrow a$. Dual, we have $a_\alpha\downarrow$ and
$a_\alpha\downarrow
 a$. A net $(a_\alpha)_{\alpha\in\varepsilon} $ is said to be order
 convergent ((o)-convergent, for short) to a point $a\in P$ if there are nets
$(u_\alpha)_{\alpha\in\varepsilon} $ and
$(v_\alpha)_{\alpha\in\varepsilon} $ of $P$ such that $a\uparrow
u_\alpha\leq a_\alpha \leq v_\alpha\downarrow a$, and we denote
$a_\alpha \xrightarrow {(o)} a$. If $\tau$ is a topology equipped on
$P$ such that every (o)-convergent net of $P$ is $\tau$- convergent,
then $\tau$ is said to have $C$ property. The strongest topology of
$P$ which has $C$ property is called the order topology of $P$ and
denoted by $\tau_o$. It is obvious that the (o)-convergence of nets
implies $\tau_o$-convergence, but the converse does not hold ([2]).
The following Theorem 2.1 answer when they are equal.

\vskip 0.1 in

A lattice $L$ is said to be (o)-continuous if $x_\alpha,x,y\in L$
and $x_\alpha\uparrow x$ implies that $x_\alpha\wedge y\uparrow
x\wedge y$.

\vskip 0.1 in

It can be proved that if $E$ is an (o)-continuous lattice effect
algebra and $x_\alpha \xrightarrow {(o)} x$ and $y_\alpha
\xrightarrow {(o)} y$, then $x_\alpha \vee y_\alpha \xrightarrow
{(o)} x\vee y$ and $x_\alpha\wedge y_\alpha \xrightarrow {(o)}
x\wedge y$ ([2]).

\vskip 0.1 in

A complete lattice effect algebra $E$ is said to be
order-topological ((o)-topological) if (o)-convergence of nets of
elements coincides with $\tau_o$-convergence and $E$ is
(o)-continuous ([2]).

\vskip 0.1 in

{\bf Lemma 2.1 ([2]).} A complete atomic (o)-continuous lattice
effect algebra $E$ is (o)-topological iff $\tau_o$ of $E$ is
Hausdorff.

\vskip 0.1 in

A partial order set $P$ is said to be down-directed if every finite
subset of $P$ has a lower bound in $P$.

\vskip 0.1 in

{\bf Lemma 2.2 ([3]).} A subset $U$ of a lattice $L$ is open in
$\tau_o$ iff for every directed subset $Y$ of $L$ and every
down-directed subset $Z$ of $L$ with $\bigvee Y=\bigwedge Z\in U$,
there exist elements $y\in Y$ and $z\in Z$ such that $[y,z]$ is
contained in $U$.

\vskip 0.1 in

An element $u$ of an effect algebra $E$ is called finite if there is
a finite sequence $\{p_1,\cdots,p_n\}$ of atoms of $E$ such that
$u=p_1\oplus\cdots \oplus p_n$. If $E$ is complete and atomic, then
for every $x\in E$ we have $x=\vee\{u\in E|u\leq x$, $u$ is
finite$\}$. Moreover, if $E$ is (o)-continuous, then the join of two
finite elements is also finite ([4]).

\vskip 0.1 in

An element $u$ of an effect algebra $E$ is called compact if $u\leq
\vee D$ for $D\subseteq E$ implies that $u\leq \vee F$ for some
finite subset $F\subseteq D$, and $E$ is called algebraic (or
compactly generated) if every $x\in E$ is a join of compact elements
of $E$ ([4]).

\vskip 0.1 in

{\bf Lemma 2.3 ([4]).} Let $E$ be a complete atomic (o)-continuous
lattice effect algebra. Then for every finite element $u$ of $E$, if
$u\leq \vee D$ for $D\subseteq E$ implies that $u\leq \vee F$ for
some finite subset $F$ of $D$.

\vskip 0.1 in

{\bf Lemma 2.4.} Let $E$ be a complete atomic (o)-continuous effect
algebra. Then for each finite element $u$, $[u,1]$ and $[0,u']$ are
$\tau_o$-clopen sets.

\vskip 0.1 in

{\bf Proof.} Evidently, $[u,1]$ and $[0,u']$ are $\tau_o$-closed
sets. Let $Y$ be a directed subset of $E$ and $Z$ be a down-directed
subset of $E$ and satisfy $\bigvee Y=\bigwedge Z\in [u,1]$. As $u$
is finite, by Lemma 2.3, there exists a subset
$\{y_1,\cdots,y_n\}\subseteq Y$ such that $u\leq \vee_{i=1}^ny_i.$
Since $Y$ is directed, there is a $y_0\in Y$ such that $u\leq
\vee_{i=1}^ny_i\leq y_0.$ For every fixed $z_0\in Z$,
$y_0\leq\bigvee Y=\bigwedge Z\leq z_0$, so $[y_0,z_0]\subseteq
[u,1]$. By Lemma 2.2, $[u,1]$ is $\tau_o$-open.

\vskip 0.1 in

Similarly, we can prove $[0,u']$ is $\tau_o$-open set.

\vskip 0.1 in

A lattice $L$ is said to be totally order-disconnected if its
lattice operations are order topology $\tau_o$ continuous and for
any two elements $x,y$ with $x\nleq y$, there exists a clopen upper
set $U$ containing $x$ but not $y$, where $U$ is an upper set iff
$u\in U$ implies $\{x\in L:x\geq u\}\subseteq U$.

\vskip 0.1 in

{\bf Lemma 2.5.} Let $E$ be a complete atomic (o)-continuous lattice
effect algebra. Then $E$ is totally order-disconnected.

\vskip 0.1 in

{\bf Proof.} Let $a,b$ be two elements of $E$ with $a\nleq b $. As
$a=\vee \{u\in E|u\leq a$, $u$ is finite\}, there exists a finite
element $u_0\in E$ such that $u_0\leq a$ and $u_0\nleq b$. Let
$U_1=[u_0,1]$. Then $a\in U_1$ and $b\notin U_1$. By Lemma 2.4,
$U_1$ is an upper set and $\tau_o$-clopen. Thus, $E$ is totally
order-disconnected.

\vskip 0.1 in

It is clear that for every totally order-disconnected lattice, its
order topology is disconnected and Hausdorff.

\vskip 0.1 in

The following theorem establishes the equivalent relation among
(o)-continuity, (o)-topological, totally order-disconnected and
algebra property of a complete atomic lattice effect algebra $E$.

\vskip 0.1 in

{\bf Theorem 2.1} Let $E$ be a complete atomic lattice effect
algebra. The following statements are equivalent:

(1) $ \ $ $E$ is order-continuous.

(2) $ \ $ $E$ is order-topological.

(3) $ \ $ $E$ is totally order-disconnected.

(4) $ \ $ $E$ is algebraic.

\vskip 0.1 in

{\bf Proof.} (1) $\Rightarrow$ (2). It follows from Lemma 2.5 that
$\tau_o$ is Hausdorff, so by Lemma 2.1, (2) holds.  (1)
$\Rightarrow$ (3) can be proved by Lemma 2.5 and (1) implies (2).
(1) $\Rightarrow$ (4) is obtained by Lemma 2.3. (3) $\Rightarrow$
(2): Since (3) implies that $\tau_o$ is Hausdorff and the binary
operations of $\wedge$ and $\vee$ are $\tau_o$ continuous, by the
similar method with the proof of Theorem 8 in [2], (2) holds. For
(4) $\Rightarrow$ (1), we refer to [3]. (2) $\Rightarrow$ (1) is
clear.

\vskip 0.2 in

{\bf  3. The relation of order topology and Frink ideal topology of
effect algebras}

\vskip 0.2 in

The Frink ideal topology is an important intrinsic topology of
partial order set. Frink pointed out that the topology is the
correct topology for chains and direct products of a finite numbers
of chains. Atherton in [5] asked: Whether the Frink ideal topology
is Hausdorff topology in every distributive lattice. Ward pointed
out that in a Boolean algebra, the Frink ideal topology is Hausdorff
topology and it is quite usual for the Frink ideal topology to be
strictly finer than the order topology ([6]). Now, we show that for
every complete atomic distributive lattice effect algebra $E$, the
Frink ideal topology $\tau_{id}$ is Hausdorff and is finer than its
order topology $\tau_{o}$, and $\tau_{id}=\tau_o$ iff $1$ is finite
iff every element of $E$ is finite iff $\tau_{id}$ and $\tau_o$ are
discrete topologies.

\vskip 0.1 in

Let $L$ be a lattice and $I\subseteq L$. Then $I$ is said to be an
ideal of $L$ if the following conditions are satisfied:

(i) $ \ $ When $a\in I$, $x\in L$ and $x\leq a$, $x\in I$.

(ii) $ \ $ When $a\in I,\ b\in I$, $a\vee b\in I$.

\vskip 0.1 in

The ideal $I$ of $L$ is said to be a completely irreducible ideal if
it is not the intersection of a collection of ideals all distinct
from it, i.e., if $(I_\alpha)_{\alpha\in \Lambda}$ is a collection
of ideals such that $I=\cap_{\alpha\in \Lambda} I_\alpha$, then
$I=I_{\alpha_0}$ for some $\alpha_0\in \Lambda$. It is clear that
every maximal ideal is a completely irreducible ideal.

\vskip 0.1 in

Similarly, the dual ideal and completely irreducible dual ideal of
$L$ can be defined, too.

\vskip 0.1 in

Let $L$ be a lattice. The Frink ideal topology $\tau_{id}$ of $L$
can be described as following: Take all completely irreducible
ideals and completely irreducible dual ideals of $L$ as a subbasis
of the open sets of the topology $\tau_{id}$ ([7]).

\vskip 0.1 in

Elements $a,b$ of a lattice effect algebra $E$ are said to be
compatible iff $a\vee b=a\oplus (b\ominus(a\wedge b))$ and denoted
by $a\leftrightarrow b$. If for any $a,b\in E$, $a\leftrightarrow
b$, then $E$ is said to be a $MV$-effect algebra ([8]).

\vskip 0.1 in

In order to prove our main results in this section, we first need
the following:

\vskip 0.1 in

{\bf Lemma 3.1 ([9]).} Let $E$ be a lattice effect algebra.

(i) $ \ $ If $x\oplus y$ exists, then $x\oplus y=(x\vee y)\oplus
(x\wedge y)$.

(ii) $ \ $ If $x\wedge y=0$ and for $m,n\in \bf N$, the elements
$mx$, $ny$ and $(mx)\oplus (ny)$ exist in $E$, then $(kx)\wedge
(ly)=0$ and $(kx)\vee(ly)=(kx)\oplus(ly)$ for all $k\in
\{1,\cdots,m\}$, $l\in \{1,\cdots,n\}$.

(iii) $ \ $ Let $Y\subseteq E$. If $\bigvee Y$ exists in $E$ and
$x\in E$ such that $x\leftrightarrow y $ for every $y\in Y$, then
$x\wedge (\bigvee Y)=\bigvee \{x\wedge y:y\in Y\}$ and
$x\leftrightarrow \bigvee Y$.

\vskip 0.1 in

A finite subset $F=(a_k)_{k=1}^n$ of effect algebra $E$ is said to
be $\oplus$-orthogonal if $a_1\oplus a_2\oplus\cdots \oplus a_n$
exists in $E$ and denote $a_1\oplus a_2\oplus\cdots \oplus a_n$ with
$\bigoplus_{k=1}^na_k$ or $\bigoplus F$. An arbitrary subset
$G=(a_k)_{k\in H}$ of $E$ is said to be $\oplus$-orthogonal if
$\bigoplus K$ exists for every finite subset $K\subseteq G$. Let
$G=(a_k)_{k\in H}$ be a $\oplus$-orthogonal subset of $E$. If
$\bigvee \{\bigoplus K|K\subseteq G$ finite$\}$ exists in $E$, we
denote $\bigvee \{\bigoplus K|K\subseteq G$ finite$\}$ with
$\bigoplus G$ ([9]).

\vskip 0.1 in

{\bf Lemma 3.2 ([9]).} Let $E$ be a complete effect algebra and
$(a_k)_{k\in H}$ a $\oplus$-orthogonal subset of $E$. If
$H_1\subseteq H$, $H_2=H\backslash H_1$, then
$$\bigoplus_{k\in H} a_k=(\bigoplus_{k\in H_1}a_k)\oplus
(\bigoplus_{k\in H_2}a_k).$$

\vskip 0.1 in

For an element $x$ of an effect algebra $E$, we define
ord$(x)=\infty$ if $nx=x\oplus\cdots\oplus x$ ($n$ times) exists for
every $n\in N$ and ord$(x)=n_x\in N$ (called isotropic index) if
$n_x$ is the greatest integer such that $n_xx$ exists in $E$. It is
clear that in a complete lattice effect algebra $E$, $n_x<\infty$
for every $x\in E$.

\vskip 0.1 in

The set of sharp elements of $E$ is denoted by $S(E)$. It has been
shown that in every lattice effect algebra $E$, $S(E)$ is an
orthomodular lattice, is a sub-effect algebra and a sublattice of
$E$ ([4]). Moreover, $S(E)$ is a full sub-lattice of $E$, that is,
$S(E)$ inherits all suprema and infima of subsets of $S(E)$ if they
exist in $E$. In a complete atomic distributive lattice effect
algebra $E$, $S(E)$ is a complete atomic Boolean algebra ([9]).

\vskip 0.1 in

An element $x\in E$ is said to be principle if $a\leq x,b\leq x$ and
$a\oplus b$ is defined, then $a\oplus b\leq x$.

\vskip 0.1 in

In a lattice effect algebra $E$, $x$ is principle iff $x$ is sharp.

\vskip 0.1 in

{\bf Lemma 3.3 ([9]).} If $E$ is an atomic lattice effect algebra
and $a\in E$ is an atom with ord$(a)=n_a$, then

(i) $ \ $ $(ka)\wedge (ka)'\neq 0$ for all $k\in \{1,2,\cdots,
n_a-1\}$.

(ii) $ \ $ $n_aa\in S(E)$.

(iii) $ \ $ If $u=(k_1a_1)\oplus (k_2a_2)\oplus\cdots \oplus
(k_na_n)$, where $\{a_1,a_2,\cdots,a_n\}$ is a set of mutually
different atoms of $E$, then $u=\vee_{i=1}^n(k_ia_i)$.

(iv) $ \ $ If $E$ is complete and $x\neq 0$, then there are mutually
different atoms $a_\alpha\in E, \alpha\in \Gamma$, and positive
integers $k_\alpha$ such that $$x=\bigoplus\{k_\alpha
a_\alpha|\alpha\in \Gamma\}=\bigvee \{k_\alpha a_\alpha|\alpha\in
\Gamma\}.$$ Moreover, $x\in S(E)$ iff $k_\alpha=n_{a_\alpha}=$
ord$(a_\alpha)$ for all $\alpha\in \Gamma$.

\vskip 0.1 in

{\bf Lemma 3.4 ([9]).} Let $E$ be a complete atomic effect algebra.
Then for every $x\in E$ with $x\neq 0$, there exists a unique
$w_x\in S(E)$, a unique set $\{a_\alpha |\alpha\in \mathcal{A}\}$ of
atoms of $E$ and unique positive integers $k_\alpha$, $\alpha\in
\mathcal{A}$, such that $$x=w_x\oplus (\bigoplus\{k_\alpha
a_\alpha|\alpha\in \mathcal{A}\}).$$ Moreover, if $w\in S(E)$ with
$w\leq x\ominus w_x$, then $w=0$.

\vskip 0.1 in

{\bf Definition 3.1 ([10-12]).} An effect algebra $E$ is called
sharply dominating if for every $a\in E$ there exists a smallest
sharp element $\hat{a}\in E$ such that $a\leq \hat{a}$. A sharply
dominating effect algebra $E$ is called $S$-dominating if $a\wedge
p$ exists for every $a\in E$ and $p\in S(E)$.

\vskip 0.1 in

It is clear that a lattice effect algebra $E$ is $S$-dominating iff
$E$ is sharply dominating. Every complete lattice effect algebra is
sharply dominating.

\vskip 0.1 in

{\bf Lemma 3.5.} Let $E$ be a complete atomic distributive lattice
effect algebra. If $u$ is a finite element of $E$, then the smallest
sharp element $\hat{u}$ dominating $u$ is also finite. If
$u_1,u_2\in E$ with $u_1\leq u_2$, then $\hat{u_1}\leq \hat{u_2}$.

\vskip 0.1 in

{\bf Proof.} By Lemma 3.3, there are mutually distinct atoms
$\{a_1,\cdots, a_m\}$ and positive integers $\{k_1,\cdots, k_m\}$
such that $u=\oplus_{i=1}^m k_ia_i=\vee_{i=1}^m k_ia_i$. We claim
$\hat{u}=\oplus_{i=1}^m n_ia_i=\vee_{i=1}^m n_ia_i$, where $n_i$ is
the isotropic index of $a_i$. Let $b\in S(E)$ with $u\leq b$. For
every $i$ with $k_i\neq n_i$, we have $k_ia_i\oplus b'$ is defined
and since $(k_ia_i)\wedge b'\leq b\wedge b'=0$, we obtain, by Lemma
3.1, that $(k_ia_i)\oplus b'=(k_ia_i)\vee b'$. As $k_ia_i\leq a_i'$
and $b'\leq (k_ia_i)'\leq a_i'$, $(k_ia_i)\oplus b'=(k_ia_i)\vee
b'\leq a_i'$. It follows that there exists $(k_i+1)a_i\oplus
b'=(k_i+1)a_i\vee b'\leq a_i'$. Hence, $(k_i+2)a_i\oplus b'$ exists,
by induction, $n_ia_i\oplus b'$ exists and so $n_ia_i\leq b$. Thus,
$\hat{u}=\vee_{i=1}^{m}n_ia_i\leq b$.

Let $u_1\leq u_2$. Then $u_1\leq u_2\leq \hat{u_2}$. By the
definition of sharply dominating effect algebra, we get
$\hat{u_1}\leq \hat{u_2}$.

\vskip 0.1 in

{\bf Lemma 3.6.}  Let $E$ be a complete atomic distributive lattice
effect algebra and $F$ the set of all finite elements and $0$ of
$E$. Then $F$ is an ideal of $E$.

\vskip 0.1 in

{\bf Proof.} Note that in every complete atomic distributive lattice
effect algebra, the join of two finite elements is finite as well,
so we only need to prove the fact that if $u\in E$ is finite and
$x\in E$ with $0\neq x\leq u$, then $x$ is finite.

Let $u$ be finite and $0\neq x\leq u$. Then, by Lemma 3.5,
$\hat{u}\in S(E)$ and $\hat{u}$ is finite with $x\leq \hat{u}$. It
follows from Lemma 2.3 that $\hat{u}$ is a compact element of $E$,
so it is a compact element of $S(E)$.

(i) If $x\in S(E)$, by Lemma 3.3, we can assume $x=\oplus_{\alpha\in
\Lambda}n_\alpha a_\alpha=\vee_{\alpha\in \Lambda}n_\alpha
a_\alpha$, where $\{a_\alpha:\alpha\in \Lambda\}$ is a set of atoms
and $n_\alpha$ is the isotropic index of $a_\alpha$. Note that
$S(E)$ is a complete atomic Boolean algebra and $x\leq \hat{u}$, so
$x$ is a compact element of $S(E)$. Thus, there exist
$\{\alpha_1,\cdots,\alpha_m\}\subseteq \Lambda $ such that
$x=\vee_{\alpha\in \Lambda} n_\alpha a_\alpha\leq \vee_{i=1}^m
n_{\alpha_i} a_{\alpha_i}$, so $x=\vee_{i=1}^m n_{\alpha_i}
a_{\alpha_i}=\oplus_{i=1}^mn_{\alpha_i} a_{\alpha_i}$. That is, $x$
is a finite element of $E$.

(ii) If $x\notin S(E)$. There exists $x_1\in E$ such that
$\hat{u}=x\oplus x_1$. By Lemma 3.4, we can assume that
$$x=w_x\oplus(\bigvee _{\alpha\in \Lambda}k_\alpha
b_\alpha)=w_x\oplus(\bigoplus _{\alpha\in \Lambda}k_\alpha
b_\alpha),$$ $$x_1=w_{x_1}\oplus(\bigvee _{\beta\in \Gamma}l_\beta
c_\beta)=w_{x_1}\oplus(\bigoplus _{\beta\in \Gamma}l_\beta
c_\beta),$$ where $w_x,w_{x_1}\in S(E)$, $\{b_\alpha:\alpha\in
\Lambda\}$ and $\{c_\beta:\beta\in \Gamma\}$ are sets of atoms and
$k_\alpha\neq n_\alpha$, $l_\beta\neq n_\beta$ for every $\alpha\in
\Lambda$ and $\beta\in \Gamma$. Note that $S(E)$ is a sub-effect
algebra, denote $x_0=(\bigoplus_{\alpha\in \Lambda}k_\alpha
b_\alpha)\oplus (\bigoplus_{\beta\in \Gamma}l_\beta c_\beta)$, we
obtain $x_0=\hat{u}\ominus w_x\ominus w_{x_1}\in S(E)$. Denote
$\Lambda_1=\{\alpha\in \Lambda:$ there exists $c_\beta$ such that
$b_\alpha=c_\beta\}$ and $\Gamma_1=\{\beta\in \Gamma:$ there exists
$b_\alpha$ such that $c_\beta=b_\alpha\}$. For every $\beta\in
\Gamma_1$, if $c_\beta=b_\alpha$, then we denote $l_\alpha=l_\beta$.
Thus, by Lemma 3.2,
$$(\bigoplus_{\alpha\in \Lambda}k_\alpha b_\alpha)\oplus(\bigoplus_{\beta\in
\Gamma}l_\beta c_\beta)=\bigoplus_{\alpha\in \Lambda_1}(k_\alpha+
l_\alpha)b_\alpha\oplus (\bigoplus_{\alpha\in {\Lambda\setminus
\Lambda_1}}k_\alpha b_\alpha)\oplus (\bigoplus_{\beta\in
{\Gamma\setminus \Gamma_1}}l_\beta c_\beta). $$ So $b_\alpha\neq
c_\beta $ for every $\alpha\in (\Lambda\setminus \Lambda_1)$ and
$\beta\in (\Gamma\setminus \Gamma_1)$. It follows from Lemma 3.1
that $(k_\alpha b_\alpha)\wedge (l_\beta c_\beta)=0$ for every
$\alpha\in (\Lambda\setminus \Lambda_1)$ and $\beta\in
(\Gamma\setminus \Gamma_1)$. As $(k_\alpha b_\alpha)\oplus (l_\beta
c_\beta)$ is defined, $(k_\alpha b_\alpha)\leftrightarrow (l_\beta
c_\beta)$, where $\alpha\in (\Lambda\setminus \Lambda_1)$ and
$\beta\in (\Gamma\setminus \Gamma_1)$. By Lemma 3.1, we have
$$k_\alpha b_\alpha\leftrightarrow \bigvee_{\beta\in\Gamma\setminus
\Gamma_1} (l_\beta c_\beta),\ \  \bigvee_{\alpha\in \Lambda
\setminus \Lambda_1}(k_\alpha b_\alpha)\leftrightarrow
\bigvee_{\beta\in\Gamma\setminus \Gamma_1} (l_\beta c_\beta).$$ So
$$(\bigvee_{\alpha\in \Lambda \setminus \Lambda_1}(k_\alpha
b_\alpha))\wedge (\bigvee_{\beta\in\Gamma\setminus \Gamma_1}
(l_\beta c_\beta))=\bigvee_{\alpha\in \Lambda \setminus
\Lambda_1,\beta\in\Gamma\setminus \Gamma_1}((k_\alpha
b_\alpha)\wedge(l_\beta c_\beta))=0.$$ Hence $$(\bigvee_{\alpha\in
\Lambda \setminus \Lambda_1}(k_\alpha b_\alpha))\oplus
(\bigvee_{\beta\in\Gamma\setminus \Gamma_1}( l_\beta
c_\beta))=(\bigvee_{\alpha\in \Lambda \setminus \Lambda_1}(k_\alpha
b_\alpha))\vee (\bigvee_{\beta\in\Gamma\setminus \Gamma_1} (l_\beta
c_\beta)).$$ That is $$ (\bigoplus_{\alpha\in {\Lambda\setminus
\Lambda_1}}k_\alpha b_\alpha)\oplus (\bigoplus_{\beta\in
{\Gamma\setminus \Gamma_1}}l_\beta c_\beta)=(\bigvee_{\alpha\in
\Lambda \setminus \Lambda_1}(k_\alpha b_\alpha))\vee
(\bigvee_{\beta\in\Gamma\setminus \Gamma_1} (l_\beta c_\beta)).$$
Similarly, we can prove $x_0=(\bigvee_{\alpha\in
\Lambda_1}(k_\alpha+ l_\alpha)b_\alpha)\vee (\bigvee_{\alpha\in
{\Lambda\setminus \Lambda_1}}k_\alpha b_\alpha)\vee
(\bigvee_{\beta\in {\Gamma\setminus \Gamma_1}}l_\beta c_\beta)$. For
every fixed $\alpha_0\in \Lambda\setminus \Lambda_1$,
$(k_{\alpha_0}a_{\alpha_0})\wedge x_0'\leq x_0\wedge x_0'=0$. As
$\bigoplus_{\alpha\in \Lambda_1}(k_\alpha+ l_\alpha)b_\alpha\oplus
(\bigoplus_{\alpha\in {\Lambda\setminus \Lambda_1}}k_\alpha
b_\alpha)\oplus (\bigoplus_{\beta\in {\Gamma\setminus
\Gamma_1}}l_\beta c_\beta)$ is defined, we have
$$k_{\alpha_0}a_{\alpha_0}\leq  (\bigvee_{\alpha\in
\Lambda_1}(k_\alpha+ l_\alpha)b_\alpha)',\
k_{\alpha_0}a_{\alpha_0}\leq (\bigvee_{\alpha\in {\Lambda\setminus
\Lambda_1},\alpha\neq \alpha_0}k_\alpha b_\alpha)',\
k_{\alpha_0}a_{\alpha_0}\leq (\bigvee_{\beta\in {\Gamma\setminus
\Gamma_1}}l_\beta c_\beta)'.$$ So $(k_{\alpha_0}a_{\alpha_0})\wedge
x_0'=(k_{\alpha_0}a_{\alpha_0})\wedge ((\bigvee_{\alpha\in
\Lambda_1}(k_\alpha+l_\alpha)b_\alpha)\vee (\bigvee_{\alpha\in
{\Lambda\setminus \Lambda_1}}k_\alpha b_\alpha)\vee
(\bigvee_{\beta\in {\Gamma\setminus \Gamma_1}}l_\beta
c_\beta))'=(k_{\alpha_0}a_{\alpha_0})\wedge
(k_{\alpha_0}a_{\alpha_0})'=0$. By Lemma 3.3, we have
$k_{\alpha_0}=n_ {\alpha_0}$. Note that we have assumed that
$k_\alpha\neq n_\alpha$ for every $\alpha\in {\Lambda }$, so
$\Lambda= \Lambda_1$. Similarly, we have $\Gamma=\Gamma_1$ and
$k_\alpha +l_\alpha=n_\alpha$ for every $\alpha\in \Lambda$. That is
$$(\bigoplus_{\alpha\in \Lambda}k_\alpha b_\alpha)\oplus
(\bigoplus_{\beta\in \Gamma}l_\beta c_\beta)=\bigvee_{\alpha\in
\Lambda}(k_\alpha+l_\alpha)b_\alpha=\bigvee_{\alpha\in
\Lambda}n_\alpha b_\alpha=\hat{u}\ominus w_x\ominus w_{x_1}\in
S(E).$$ Since $\bigvee_{\alpha\in \Lambda}n_\alpha b_\alpha\leq
\hat{u}$, $\hat{u}$ is a compact element of $S(E)$ and $S(E)$ is a
complete atomic Boolean algebra, we get $\bigvee_{\alpha\in
\Lambda}n_\alpha b_\alpha$ is compact. Thus, there exists
$\{\alpha_1,\cdots,\alpha_m\}\subseteq \Lambda$ such that
$$\bigvee_{\alpha\in \Lambda}n_\alpha b_\alpha=\bigvee_{i=1}^m
n_{\alpha_i}b_{\alpha_i}=\bigoplus_{i=1}^m
n_{\alpha_i}b_{\alpha_i}.$$ As $\bigvee_{\alpha\in \Lambda}n_\alpha
b_\alpha= \bigoplus_{\alpha\in \Lambda}n_\alpha
b_\alpha=\bigoplus_{i=1}^m n_{\alpha_i}b_{\alpha_i}$, $\Lambda$ is
finite. Therefore $x=w_x\oplus(\bigvee_{\alpha\in \Lambda}k_\alpha
b_\alpha)$ is finite. The lemma is proved.

\vskip 0.1 in

Recall that the interval topology $\tau_i$ of an effect algebra $E$
is the topology which is defined by taking all closed interval
$[a,b]$ as a sub-basis of closed sets of $E$. It is well known that
$\tau_{id}\geq \tau_i$ and $\tau_{o}\geq \tau_i$ in every lattice.

\vskip 0.1 in

{\bf Lemma 3.7 ([2]).} Let $E$ be a complete atomic distributive
lattice effect algebra. Then its interval topology $\tau_i$ is
compact Hausdorff topology and $\tau_o=\tau_i$.

\vskip 0.1 in

{\bf Theorem 3.1.}  Let $E$ be a complete atomic distributive
lattice effect algebra. Then $\tau_{id}$ is Hausdorff and
$\tau_{id}\geq \tau_o$. Moreover, the following conditions are
equivalent:

(i) $ \ $ $\tau_{id}=\tau_o$.

(ii) $ \ $ $1$ is finite.

(iii)  $ \ $ Every element of $E$ is finite.

(iv)  $ \ $ $\tau_{id}$ and $\tau_o$ are both discrete topologies.

\vskip 0.1 in

{\bf Proof.} It follows from Lemma 3.7 and $\tau_{id}\geq \tau_i$
that $\tau_{id}$ is Hausdorff topology and $\tau_{id}\geq \tau_o$.

(ii) $\Leftrightarrow $ (iii) can be proved by Lemma 3.6 easily.

(iii) $\Rightarrow $ (iv). For every $x\in E$, $x$ and $x'$ are both
finite. By Lemma 2.4, $[0,x]$ and $[x,1]$ are $\tau_o$-open, so
$\{x\}=[0,x]\cap [x,1]$ is $\tau_o$ open, this showed that $\tau_o$
is discrete. Note that $\tau_{id}\geq \tau_o$, we have $\tau_{id}$
and $\tau_o$ are both discrete.

(iv) $\Rightarrow $ (i) is obvious.

(i) $\Rightarrow $ (ii).  Assume $1$ is not finite. Let $F_0$ be the
set of all finite elements and $0$. Then it follows from Lemma 3.6
that $F_0$ is an ideal and $1\notin F_0$. By the Zorn's Lemma, $F_0$
is contained in an ideal $F$ maximal subject to not containing $1$.
It is easy to prove that $F'=\{f'\in E:f\in F\}$ is a maximal dual
ideal and so $\tau_{id}$-open with $1\in F'$. As $1=\vee \{u\in E:u$
is finite\}, we can choose a net of finite elements
$(u_\alpha)_{\alpha\in \Lambda}$ of $E$ such that $u_\alpha \uparrow
1$, thus, by Lemma 3.5, $(\hat{u}_\alpha)_{\alpha\in \Lambda}$ is
also a net of finite elements of $E$ and $\hat{u}_\alpha \uparrow
1$. Note that $(u_\alpha)_{\alpha\in \Lambda} \subseteq F$ and
$\hat{u}_\alpha \vee \hat{u}_\alpha '=1$, we have $\hat{u}_\alpha
'\notin F$, otherwise $1\in F$. Hence $\hat{u}_\alpha \notin F'$.
That is, $\hat{u}_\alpha$ is not $\tau_{id}$-convergent to $1$.
However, $\hat{u}_\alpha\xrightarrow{(\tau_{o})} 1$. This
contradicts (i). So $1$ is finite.

\vskip 0.2 in

{\bf 4. The order topology continuity of operation $\oplus$ of
effect algebras}

\vskip 0.2 in

For the order topology continuity of operation $\oplus$, Wu had only
presented a sufficient condition under a very strictly assumption
([13]). Now, we study this question continuously.

\vskip 0.1 in

{\bf Lemma 4.1 ([14]).} Let $(X,T)$ be a topology space. Then
$(X,T)$ is Hausdorff space iff every convergent net in $X$ has
exactly one limit point.

\vskip 0.1 in

{\bf Lemma 4.2 ([2]).} Let $E$ be a complete (o)-continuous lattice
effect algebra, $x_\alpha,x,y\in E$. Then

(i) $ \ $ $x_\alpha\xrightarrow {(\tau_o)} x\Rightarrow x_\alpha\vee
y\xrightarrow {(\tau_o)} x\vee y$.

(ii) $ \ $  $x_\alpha\xrightarrow {(\tau_o)} x\Rightarrow
x_\alpha\wedge y\xrightarrow {(\tau_o)} x\wedge y$.

(iii) $ \ $ $x_\alpha\xrightarrow {(\tau_o)} x\Rightarrow x_\alpha
'\xrightarrow {(\tau_o)} x'$.

\vskip 0.1 in

Our main results in this section are the following:

\vskip 0.1 in

{\bf Theorem 4.1.} Let $(E,\oplus,0,1)$ be a complete (o)-continuous
lattice effect algebra. If $\oplus$ is order topology $\tau_o$
continuous, then $\tau_o$ is Hausdorff topology.

\vskip 0.1 in

{\bf Proof.} Let $(x_\alpha)_{\alpha\in \Lambda}$ be a net in $E$
and $x_\alpha\xrightarrow{(\tau_o)} x$,
$x_\alpha\xrightarrow{(\tau_o)} y$. By Lemma 4.2, $$x_\alpha\vee
x\xrightarrow{(\tau_o)}  x,\ x_\alpha \vee x\xrightarrow{(\tau_o)}
x\vee y,\ (x_\alpha \vee x)'\xrightarrow{(\tau_o)} (x\vee y)'.$$It
follows from $\oplus$ is $\tau_o$ continuous that
$$(x_\alpha\vee x)\oplus (x_\alpha\vee x)'\xrightarrow{(\tau_o)}
x\oplus (x\vee y)'.$$ That is, $1\xrightarrow{(\tau_o)} x\oplus
(x\vee y)'$. Note that $\tau_o\geq \tau_i$, $1\xrightarrow{(\tau_i)}
x\oplus (x\vee y)'$ and $\{1\}$ is $\tau_i$-closed, we have $x\oplus
(x\vee y)'=1$, so $x=x\vee y$, $y\leq x$. Similarly, we can prove
that $x\leq y$. Thus $x=y$ and $\tau_o$ is Hausdorff topology.

\vskip 0.1 in

{\bf Example 4.1.} Let $L$ be the complete Boolean algebra of all
regular open subsets of the unit interval $I$. It follows from $L$
is (o)-continuous and the order topology $\tau_o$ of $L$ is not
Hausdorff topology ([15]) and Theorem 4.1 that $\oplus$ is not order
topology $\tau_o$ continuous.

\vskip 0.1 in

{\bf Theorem 4.2.} Let $(E,\oplus,0,1)$ be a complete atomic
(o)-continuous lattice effect algebra. If
$a_\alpha\xrightarrow{(\tau_o)} 0$ and
$b_\alpha\xrightarrow{(\tau_o)} 0$ with $a_\alpha\leq b_\alpha '$
for every $\alpha$, then $a_\alpha\oplus b_\alpha
\xrightarrow{(\tau_o)} 0$.

\vskip 0.1 in

{\bf Proof.} Suppose $\wedge_{\beta}\vee_{\alpha\geq
\beta}(a_\alpha\oplus b_\alpha)=c$. For every finite element $u\in
E$, there exists sharply dominating element $\hat{u}$ such that
$u\leq \hat{u}$. It follows from Lemma 3.5 that $\hat{u}$ is also
finite. So $[0,\hat{u}']\subseteq [0,u']$ and by Lemma 2.4 that they
are both $\tau_o$-open. Note that $0\in [0,\hat{u}']$, there exists
$\alpha_0$ such that for every $\alpha\geq \alpha_0$, $a_\alpha\in
[0,\hat{u}']$ and $b_\alpha\in [0,\hat{u}']$. Thus, for every
$\alpha\geq \alpha_0$, $a_\alpha\oplus b_\alpha\leq \hat{u}'$ since
$\hat{u}'$ is principle, so $\vee_{\alpha\geq
\alpha_0}(a_\alpha\oplus b_\alpha)\leq \hat{u}'$ and $c\leq
\hat{u}'$, this showed that $c'\geq \hat{u}\geq u$. As $1=\vee\{u\in
E: u$ is finite $\}$, $c'\geq 1$. That is, $c'=1$ and $c=0$. Hence,
we have $a_\alpha\oplus b_\alpha \xrightarrow{(o)} 0$, thus,
$a_\alpha\oplus b_\alpha \xrightarrow{(\tau_o)} 0$.

\vskip 0.1 in

{\bf Lemma 4.3 ([16]).} Let $E$ be a lattice effect algebra. Then
$x_\alpha\xrightarrow{(o)} x$ iff $(x\vee x_\alpha)\ominus
x\xrightarrow{(o)}0$ and $x\ominus (x\wedge x_\alpha)
\xrightarrow{(o)}0$.

\vskip 0.1 in

{\bf Theorem 4.3.} Let $(E,\oplus,0,1)$ be a complete atomic
(o)-continuous lattice effect algebra. If
$a_\alpha\xrightarrow{(\tau_o)} a$ and
$b_\alpha\xrightarrow{(\tau_o)} b$ with $a_\alpha\leq b_\alpha '$
and $a_\alpha\leq b'$ and $b_\alpha\leq a'$ for every $\alpha$, then
$a_\alpha\oplus b_\alpha \xrightarrow{(\tau_o)} a\oplus b$.

\vskip 0.1 in

{\bf Proof.} By Theorem 2.1 and Lemma 4.3, we only need to prove
that $$((a_\alpha\oplus b_\alpha)\vee(a\oplus b))\ominus(a\oplus b)
\xrightarrow{(\tau_o)} 0,\ (a\oplus b)\ominus ((a\oplus
b)\wedge(a_\alpha\oplus b_\alpha))\xrightarrow{(\tau_o)} 0.$$

Note that $a_\alpha\xrightarrow{(\tau_o)} a$ implies that
$a_\alpha\xrightarrow{(\tau_i)} a$, and since $a_\alpha\leq b'$ for
every $\alpha$, we have $a\leq b'$. Moreover, since $a_\alpha\leq
(b_\alpha ' \wedge b')$ and $a\leq (b_\alpha ' \wedge b')$,
$(a_\alpha \vee a)\oplus (b_\alpha \vee b)$ is defined. As
$a_\alpha\xrightarrow{(\tau_o)} a$ and
$b_\alpha\xrightarrow{(\tau_o)} b$, it follows from Lemma 4.3 that
$$(a_\alpha \vee a)\ominus a \xrightarrow{(\tau_o)} 0,\ (b_\alpha
\vee b)\ominus b \xrightarrow{(\tau_o)} 0.$$ Note that $((a_\alpha
\vee a)\ominus a)\oplus ((b_\alpha \vee b)\ominus b)=((a_\alpha \vee
a)\oplus (b_\alpha \vee b))\ominus (a\oplus b)\geq ((a_\alpha\oplus
b_\alpha)\vee(a\oplus b))\ominus(a\oplus b)$ and Theorem 4.2, we
have
$$((a_\alpha \vee a)\ominus a)\oplus ((b_\alpha \vee b)\ominus
b)\xrightarrow{(\tau_o)} 0,$$ so $$((a_\alpha\oplus
b_\alpha)\vee(a\oplus b))\ominus(a\oplus b)\xrightarrow{(\tau_o)}
0.$$ Similarly, we can prove that $(a\oplus b)\ominus ((a\oplus
b)\wedge(a_\alpha\oplus b_\alpha))\xrightarrow{(\tau_o)} 0$ and the
theorem is proved.

\vskip 0.1 in

{\bf Corollary 4.1.} Let $(E,\oplus,0,1)$ be a complete atomic
$MV$-effect algebra, $a_\alpha\xrightarrow{(\tau_o)} a$ and
$b_\alpha\xrightarrow{(\tau_o)} b$ with $a_\alpha\leq b_\alpha'$ for
every $\alpha$. Then $a_\alpha\oplus b_\alpha \xrightarrow{(\tau_o)}
a\oplus b$.

\vskip 0.2 in

{\bf Acknowledgement.} The authors wish to express their thanks to
the referees for their valuable comments and suggestions.

\end{document}